\documentclass[twocolumn]{autart}    

\usepackage{amsmath}
\usepackage{amssymb}
\usepackage{amsfonts}
\usepackage{mathrsfs}
\usepackage{color}
\usepackage{graphicx}
\usepackage{verbatim}
\usepackage{algorithm,algpseudocode,amsmath}
\usepackage{algorithmicx}
\usepackage{subfigure}
\usepackage{psfrag}
\usepackage{tabularx}
\usepackage{bm}
\usepackage{lipsum}
\usepackage{float}
\usepackage{epic,eepic}
\usepackage{amscd}
\usepackage{fancyhdr}
\usepackage{palatino}
\usepackage{shadow}
\usepackage{tikz}
\usepackage{pgf}
\usepackage{wrapfig}
\usepackage{rotating}
\usepackage[roman]{sublabel}
\usepackage{ifthen}
\usepackage{url}

\newcommand{\R}{\mathbb{R}}

\definecolor{jetting}{RGB}{126.464,47.104,142.336}
\definecolor{idle}{RGB}{0,114.432,189.696}
\definecolor{heating}{RGB}{162.56,19.968,47.104}
\definecolor{lumped}{RGB}{255,191,191}
\definecolor{pde}{RGB}{240,222,204}
\definecolor{return}{RGB}{64,255,255}
\definecolor{sol}{RGB}{0,249,0}
\definecolor{liq}{RGB}{255,64,255}

\newcommand*\cvhrulefill[2]{%
	\leavevmode\color{#1}\leaders\hrule\@height#2\hfill \kern\z@\normalcolor}
\newcommand*\crule[3]{%
	\color{#1}\rule{#2}{#3}\normalcolor}

\newcommand{\mcl}[1]{\mathcal{ #1}}

\begin{document}

\begin{frontmatter}

\title{Oscillations in Mixed-Feedback Systems\thanksref{footnoteinfo}} 

\thanks[footnoteinfo]{Corresponding author: Amritam Das}

\author{Amritam Das}\ead{ad2079@cam.ac.uk},    
\author{Thomas Chaffey}\ead{tlc37@cam.ac.uk},               
\author{Rodolphe Sepulchre}\ead{rs771@cam.ac.uk},  
     \address{University of Cambridge, Department of Engineering, Trumpington Street, Cambridge CB2 1PZ}                              


\begin{abstract}                          
A new method is presented for the analysis of limit cycle oscillations in mixed-feedback systems. The calculation
of the limit cycle is reformulated as the zero finding of a mixed-monotone relation, that is, of the difference of two maximally monotone relations. The problem can then be solved efficiently by
borrowing existing algorithms that minimize the difference of two convex functions. 
The potential of the method is illustrated on the classical 
Van der Pol oscillator.
\end{abstract}

\end{frontmatter}

\section{Introduction}

Mixed-feedback amplification is a ubiquitous mechanism in engineering and biology
to generate and control robust oscillations \cite{Sepulchre2018}.   The feedback oscillation results
from a balance between  (local) positive feedback, a  mechanism that destabilises an otherwise stable equilibrium, and (global) negative feedback, a 
mechanism that guarantees boundedness
of the solutions. The physical  interpretation is that the oscillation results from an energy balance
between locally active and globally dissipative elements. This energy balance is conveniently
expressed in the language of dissipativity theory \cite{SEPULCHRE2005809}, \cite{stan}.

The mixed-feedback mechanism has long been acknowledged in simple models like the Van der Pol oscillator,
where the desired energy balance is achieved by inserting a negative resistance element in a passive RLC circuit \cite{Casaleiro2019}.  
The same mechanism is at the core of the biophysical Hodgkin-Huxley model
of neuronal excitability \cite{hodgkin}, where the oscillation is controlled by the balance between two distinct ion channels, 
one of them providing negative conductance (see \cite{Sepulchrebook2018}).

Despite being intuitive, the mathematical analysis of oscillations in mixed-feedback systems still remains challenging
beyond the phase portrait analysis of two-dimensional models. Describing function analysis remains
the tool of choice when the oscillation is nearly harmonic (see \cite{slotine}, Chapter 5). The limitations of this approximate tool
are well understood, and it is, for instance, known to be of limited  use in the analysis of  relaxation oscillations. 
Successful alternative methods have been developed for  relay feedback systems \cite{astrom}. Relay feedback
systems can be regarded as mixed-feedback systems, where the hysteresis of the relay models the positive feedback loop. 
However, a general theory of oscillations in  mixed-feedback systems is still lacking.

The present paper explores an {\it algorithmic} angle of attack, grounded in the theory of maximal monotone relations.
The key proposal is to model a mixed-feedback system as a mixed feedback loop  of monotone relations.
The negative feedback loop preserves monotonicity, whereas the positive feedback loop locally destroys monotonicity.
In recent work, we have explored maximal monotonicity to algorithmically compute the input-output solutions of monotone relations \cite{Chaffey2020}. 
We follow the same approach here, but extend the algorithm from monotone to mixed-monotone relations. This extension
is not new in the field of optimization, and efficient algorithms have been proposed for the minimization of difference of
convex functions \cite{Horst1999}, \cite{QUYNHPHONG199473}. Such algorithms are directly applicable to the question of the present paper. We illustrate
the success of that bridge on the classical model of  the Van der Pol oscillator. 

The rest of the paper is organized as follows. After some preliminary notations and definitions in Section II, Section III presents the input-output configuration of mixed-feedback systems. In Section IV, the problem of computing the output of a mixed-feedback system for a given input is formulated as a mixed-monotone inclusion problem. Followed by a summary of the recent result on computing a monotone feedback system's output in Section V, Section VI presents the algorithmic solution of the mixed-monotone inclusion problem by borrowing theories of optimization involving the difference between two convex functions. In Section VII, the algorithm is illustrated on the Van der Pol oscillator. Finally, Section VIII presents some discussions on the developed algorithm and directions of future research.

\section{Preliminaries}

\subsubsection{Notations} The gradient of a function $f:\mathbb{R}^n\rightarrow \mathbb{R}$ is denoted by $\nabla f: \mathbb{R}^n\rightarrow \mathbb R^n$. The space of $T$-periodic, square-summable sequences in $\R^{n}$ is denoted by $l_{2,T}$. For any $u_1, u_2 \in l_{2,T}$, the inner product is denoted by $\langle u_1, u_2 \rangle:= \sum_{n = 1}^T u_1^{\top}(n)u_2(n)$ and the norm of $u\in l_{2,T}$ is denoted by $\mid\mid u \mid\mid:=\sqrt{\sum_{n = 1}^T u^{\top}(n)u(n)}$. 
\subsubsection{Convex Function} A function $f:\mathbb{R}^n\rightarrow \mathbb{R}$ is convex if for all $x_1, x_2 \in \mathbb{R}^n$, $\beta \in (0,1)$, the following inequality holds:
\[
f(\beta x_1+(1-\beta)x_2)\leq \beta f(x_1)+ (1-\beta)f(x_2).
\]
Strictness in the above inequality implies strict convexity.  
\subsubsection{Relation}
 A relation on a space $X$ is a subset $S \subseteq X\times X$.
 \begin{itemize}
 	\item We write $y \in S(u)$ to denote $(u, y) \in S$. 
 	\item The usual operations on functions can be extended to relations:
 	\begin{align*}
 	S^{-1} :=& \{(u, y)\mid y\in S(u)\}\\
 	 S+R :=& \{(u, y+z)\mid (y, u)\in S(u), (z,u) \in R\}\\
 	 SR:=& \{(x, z)\mid \exists\ y \text{ such that } (x, y) \in R, (y,z) \in S\}.
 	\end{align*}
 	\item The inverse $(\cdot)^{-1}$, in this paper, is always associated with the relational inverse unless stated otherwise. Note that, unlike the functional inverse, the relational inverse always exists , and if the functional inverse does exists, the two inverses coincide. However, given a relation $S$, in general, $SS^{-1}\neq I$ where $I:=\{(x, x)\mid x\in X\}$. 
 	\end{itemize}
\subsubsection{Monotone Relation} Let $\mathcal{H}$ be a Hilbert space, equipped with an inner product
$\langle \cdot, \cdot\rangle: \mathcal{H}\times\mathcal{H} \to \mathbb{R}$ and a norm $\mid\mid\cdot\mid\mid$.   A relation $R: \mathcal{H} \to \mathcal{H}$ is called monotone if,
for all $u_1, u_2 \in \mathcal{H}$,
\begin{align}
	\Big\langle u_1-u_2, R(u_1)-R(u_2)\Big\rangle  \geq 0.
\end{align}
 \begin{itemize}
 \item Note that this definition refers to monotonicity in the operator theoretic sense, and this is distinct from the notion of monotonicity in the sense of partial order preservation by a state-space system (cf. \cite{angeli}).
\item In classical input-output literature, montonicity on $L_2$ is known as incremental positivity. For causal relations, this is equivalent to incremental passivity. See \cite{desoer} for more details. 
 \end{itemize}
\subsubsection{Maximal Monotone Relation}  A monotone relation $R: \mathcal{H} \to \mathcal{H}$ is called \emph{maximal} if its graph $\{u, y\,|\, y =
R(u)\}$ is not properly contained within the graph of any other monotone
relation.
\begin{itemize}
\item Any continuous monotone relation is maximal.
\item The (sub)gradient of a convex and proper function is a monotone operator and if the same function is also closed then its gradient (subgradient) is maximal monotone \cite{Ryu2020}(pp. 6, 28).
 \end{itemize}

\subsubsection{Lipschitz Relation}
A relation $S:\mathcal{H}\rightarrow \mathcal{H}$ has a Lipschitz constant of $L>0$ if, for all $(u, w), (v, y) \in S$, 
\begin{align}
	\mid \mid u-v \mid \mid \leq L	\mid \mid w-y \mid \mid 
	\end{align}
\begin{itemize}
	\item If $L<1$, $S$ is called a contraction map.
	\end{itemize}
\subsubsection{Coercive and Coccoercive Relation} 
The relation $R: \mathcal{H} \to \mathcal{H}$  is said to be $\alpha$-coercive if, for all $u_1, u_2 \in \mathcal{H}$, there exists an $\alpha>0$ such that
\begin{align}
		\Big\langle u_1-u_2, R(u_1)-R(u_2)\Big\rangle  \geq \alpha \mid\mid u_1-u_2\mid\mid^{2}.
	\end{align}
The relation $R$ is called $\beta$-cocoercive for all $u_1, u_2 \in \mathcal{H}$, there exists a $\beta>0$ such that
\begin{align}
	\Big\langle u_1-u_2, R(u_1)-R(u_2)\Big\rangle  \geq \beta \mid\mid R(u_1)-R(u_2)\mid\mid^{2}.
\end{align}
\begin{itemize}
\item For causal relations, coercivity is equivalent to input-strict passivity and cocoercivity is equivalent to output-strict passivity.

\item If $A$ is $\alpha$-coercive, then $A^{-1}$ is $\frac{1}{\alpha}$-cocoercive. If $B$ is $\beta$-cocoercive, then it follows from the Cauchy-Schwarz inequality that $B$ is $\frac{1}{\beta}$-Lipschitz (for details, see \cite{Ryu2020}, section 2.2).
 \end{itemize}
\section{Input-Output Relation of Mixed-Feedback Systems}
\begin{figure}[H]
	\centering
	\includegraphics[width=0.85\linewidth]{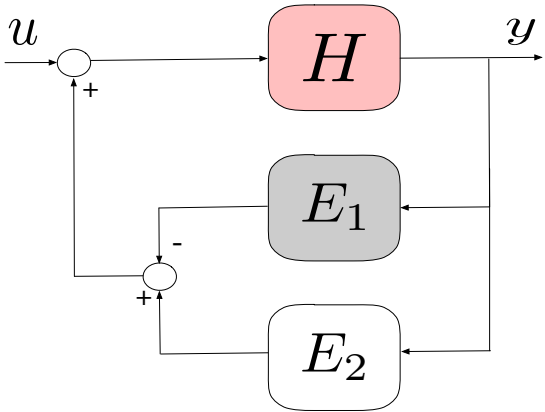}
	\caption{Input-output configuration of mixed-feeback system}
	\label{fig:mixed_feed_sec2}
\end{figure}

A mixed-feedback system is defined as a mixed feedback loop around monotone relations. Its input-output mapping is depicted as a signal-flow graph in Figure \ref{fig:mixed_feed_sec2}. Such a mixed-feedback system maps inputs signals $u\in \mathcal{H}_u$ to output signals  $y\in \mathcal{H}_y$. At this stage, the input and output spaces  $\mathcal{H}_u$ and $\mathcal{H}_y$ are considered to be Hilbert spaces. For example, in order to compute periodic solutions of mixed feedback systems, the input and output spaces can be chosen as  the spaces of $T$-periodic, square-summable sequences $l_{2,T}$. We consider $H: \mathcal{H}_u\rightarrow \mathcal{H}_y$ and $E_1, E_2: \mathcal{H}_y\rightarrow \mathcal{H}_u$ to be maximal monotone relations. Moreover, in this paper, we assume that $H$ is Linear Time Invariant (LTI). Hence, for a given $H$ (e.g. a transfer function) one may easily verify the passivity condition and consequently infer its maximal monotonicity (see  \cite{Camlibel2013}, \cite{Chaffey2020} for more details). 

We note that the forward relation can even be chosen as the identity without any loss of generality. The chosen representation is however preferred as physical examples of mixed-feedback systems typically define an input-output relation between the two terminals of a port, such as currents and voltages or forces and velocities. For instance, see figure \ref{fig:mixed_feed_vdr} for an illustration of the Van der Pol electrical circuit.

The main focus of this paper is devoted to the question of algorithmically determining the existence of a limit cycle solution in the mixed feedback system represented in figure \ref{fig:mixed_feed_sec2}.  In such case, we consider zero input or periodic input $u$ and look for a periodic output $y$ as a solution of the feedback system.

The most classical approach to answer this question would be to perform describing function analysis (c.f. \cite{hsu}, \cite{Fendrich} and the references therein). Describing function analysis provides a graphical and analytic, nevertheless \emph{approximate}, method to predict  limit cycles of nonlinear systems. Both the merits and limitations of describing function analysis are well understood \cite{slotine}. The aim of this paper is to complement the describing function analysis with an algorithmic approach as rooted in the theory of convex optimization.

\section{Problem Formulation}

The key proposal in this paper is to reformulate the existence of a limit cycle as the zero finding of a specific relation. 

To this end,  the input and output spaces are chosen as  the spaces of $T$-periodic, square-summable sequences $l_{2,T}$. We rewrite the input-output relation of mixed-feedback system in figure \ref{fig:mixed_feed_sec2}  as 
\begin{align}
	\label{io_rel}
	u \in H^{-1} y+E_1(y)-E_2(y).
\end{align}
Therefore, for a given input $u^{\star} \in l_{2,T}$, the problem of determining the output from \eqref{io_rel} amounts to computing $y = y^{\star}\in l_{2,T}$ such that 
\begin{align}
	\label{mm}
 &0\in  A(y^{\star})- B(y^{\star}),	\notag\\
	\text{with,  } & A(y) := H^{-1} y + E_1(y)-u^{\star},\  B(y) :=  E_2(y).	
\end{align}
Here, both $A$ and $ B$ are maximal monotone relations. Since, $H$ is maximal monotone, $H^{-1}$
is also maximal monotone, as maximal monotonicity is preserved by the relational inversion. The main observation from \eqref{mm} is that the mixed-feedback structure translates into the difference of two maximal monotone relations. Therefore, finding the output $y$ requires computing the zero of a \emph{mixed-motonone} relation, termed as \emph{mixed-monotone zero finding}. 

The remainder of this paper is dedicated to make use of efficient algorithms that are developed in the field of optimization and exploit them to solve the mixed-monotone zero finding problem.

\section{Algorithmic Analysis of Monotone Feedback Systems}
The property of maximal monotonicity first arose in the
study of networks of nonlinear resistors \cite{Minty1960}, \cite{Minty1961}, and has grown
to become a fundamental property in the theory of optimization
\cite{Rockafellar1976}, \cite{Parikh2013}, \cite{Ryu2020}.  Early connections between maximal
monotonicity and passive LTI systems are found in the literature on nonsmooth
dynamical systems \cite{Camlibel2016}, \cite{Brogliato2004}, and fixed point methods are
used to compute periodic solutions in nonsmooth Lur'e systems \cite{Heemels2017}.  
Recently, In \cite{Chaffey2020}, it is shown that the periodic outputs of
periodically forced maximal monotone input-output relations, built from
port connections of basic circuit elements, can be computed using fixed point
methods. The main idea of \cite{Chaffey2020} is as follows.

Given a maximal monotone relation $R: l_{2, T} \to l_{2, T}$, we can compute a solution $x
\in l_{2, T}$ to the problem
\begin{equation}
	\label{eq:zero}
	0 \in R(x),
\end{equation}

by solving for a fixed point $x_f \in
l_{2, T}$ of a related relation $F$:
\begin{align}
	x_f = F(x_f).
\end{align}
With a suitable choice of $F$, $x_f$ is a solution to \eqref{eq:zero}.  The fixed
point iteration, or Picard iteration, is given by the update rule
\begin{align}
	x^{k + 1} = F(x^{k}).
\end{align}
If $F$ is a contractive or averaged relation, the Picard iteration converges to a
fixed point of $F$, by the Banach fixed point theorem \cite{Banach1922} or the
Krasnosel'ski\u{i}-Mann theorem \cite[Theorem 5.14]{Bauschke2011}, respectively. Given an relation $R$, there is a wide range of choices for the fixed point relation
$F$, depending on the particular properties of $R$ (see \cite{Ryu2020}, \cite{Parikh2013}).

In \eqref{io_rel}, assume that the positive feedback is not present (i.e. $E_2(y) = 0$ for all $y$). Therefore, the input-output relation is now given by a maximal monotone relation, i.e. $u = H^{-1}y + E_1(y)$. Hence, finding the output $y^{\star} \in l_{2, T}$ given a particular input $u^\star \in l_{2, T}$ requires finding the zero of a monotone relation as follows:
\begin{align}
	\label{prob_monotone}
 0\in  H^{-1}y^{\star} + E_1(y^{\star}) -u^{\star}.	
\end{align}
Note that $u^\star$ is an offset, which does not affect monotonicity, and maximal monotonicity is preserved under relational summation \cite{Ryu2020}. Hence, \eqref{prob_monotone} can be solved using a fixed point iteration (see \cite{Ryu2020}, Chapter 2). In \cite{Chaffey2020}, this approach is used to compute the
periodic output of a periodically forced port circuit, where $u^\star$ represents an
input current or voltage.  

\section{Connection between Mixed-Monotone Zero-Finding Problem and DC Programing}
The theory of  maximal monotone relations has a close connection with the theory of convex optimization. In fact, for minimizing  a closed, convex, and proper function  $F(x)$, the first-order optimality condition is to find a zero of its (sub)gradient, $\nabla F(x)$, which is a maximal monotone relation \cite{Ryu2020}. In turn, if the objective is the difference of two convex functions, say $F(x)-G(x)$ where $G(x)$ is an another closed, convex, and proper function, then the first-order optimality condition is to find a zero of $\nabla F(x)- \nabla G(x)$ which is the difference of two maximal monotone relations. This correspondence suggests that methods for optimizing the difference of two convex functions can be utilised to find a zero of the difference of two maximal monotone relations, which are not necessarily the (sub)gradients of closed, convex, and proper functions.



\subsection{Difference of Convex (DC) Programing}
In literature, optimizing  the difference of convex functions is known as Difference of Convex (DC) programing.
\begin{defn}(Unconstrained DC Programing)\label{d1}
	Let $F:\mathbb{R}^n\rightarrow \mathbb{R}$ and $G:\mathbb{R}^n\rightarrow \mathbb{R}$ are closed, strictly convex, and proper functions. A Difference of Convex (DC) programing amounts to finding a local minimum of $F-G$.
\end{defn}

DC programing is a natural way to make use of convex programing in non-convex optimization. The reader is referred to \cite{Horst1999} and the references therein for an overview of DC programing and its potential applications.  Although based on heuristics, it has been used with success in a number of structured problems that include minimizing quadratic function with indefinite weight, polynomial functions etc. In fact, any function whose second partial derivatives are continuous everywhere can be expressed as the DC functions and, hence, its minimization can be cast as a DC program. Minimization of DC functions may also include constraints that are  rewritten in terms of inequalities on DC functions. 
\subsection{Disciplined DC Programing}
Particulary relevant to the present paper is the disciplined DC programing approach recommended  in \cite{shen2016disciplined}. The reasoning is as follows:  if $G$ is an affine function in definition \ref{d1}, then DC programing is equivalent to convex programing and can be solved efficiently. Hence, one way to solve a DC program and find a local optimum value is to iteratively linearize the concave function $-G$, and solve a locally convex problem. The corresponding iterative procedure is known as Convex-Concave Procedure (CCP) \cite{ccp} and can be implemented using the following algorithm.
\begin{algorithm}[H]
	\caption{: CCP for Unconstrained DC Program}
	\begin{algorithmic}[1]
		\State Initialize: $i:=0$
		\State Given: An initial guess $x_0$, $\epsilon_1$ 
		\For{$i=i+1$}
		\State Linearize $G(x)$:  $$\hat{G}(x; x_i) = G(x_i)+ \nabla G(x_i)^{\top}(x-x_i) $$
		\State Locally solve convex optimization problem: $$x_{i+1} = \arg\min\limits_{x} F(x)- \hat{G}(x; x_i)$$
		\EndFor \hspace{1ex} when
		$\frac{\max \mid x_{i+1}-x_i \mid}{\max\mid x_i\mid}< \epsilon_1$ with $x^{\star} = x_{i+1}$
		\State \textbf{return} $x^{\star}$
	\end{algorithmic}
	\label{algo_ccp}
\end{algorithm}
In Algorithm \ref{algo_ccp}, it has been proven that all the iterates are feasible and the objective value locally converges \cite{Lipp}.   

\subsection{Solving Mixed-Monotone Inclusion Problems Using DC Programing}
Disciplined DC programing provides a clear path to an iterative algorithm that computes a zero of the difference of the monotone operator $A(x) - B(x)$. Here, linearizing the concave term at each iteration in Algorithm 1 simply translates into evaluating the non-monotone term at each iteration, since $A(x)-B(a)$ is monotone for each constant $a$. This leads to the following iteration:
	
	\begin{algorithm}[H]
		\caption{: Mixed-Monotone Zero Finding}
		\begin{algorithmic}[1]
			\State Initialize: $i:=0$
			\State Given: An initial guess $x_0$, $\epsilon_1$
			\For{$i=i+1$}
			\State Solve for $x_{i+1}$ given:  $$0\in  A(x_{i+1})- B(x_i) $$
			\EndFor \hspace{1ex} when
			$\frac{\max \mid x_{i+1}-x_i \mid}{\max\mid x_i\mid}< \epsilon_1$ with $x^{\star} = x_{i+1}$
			\State \textbf{return} $x^{\star}$
		\end{algorithmic}
		\label{algo_mixed_inclusion}
	\end{algorithm}
\begin{rem} (Reasoning Behind Algorithm \ref{algo_mixed_inclusion})
	
If $x^{\star}$ is a root of $A(x) - B(x)$, then $A(x) - B(x^{\star})$
	clearly has $ x = x^{\star}$ as a root. The point $x^{\star}$ is therefore a
	fixed point of the scheme above. The converse
	statement also holds true.
%
\end{rem}

Algorithm \ref{algo_mixed_inclusion} provides an iterative approach for solving the {\it mixed-monotone} zero finding problem by solving a sequence of {\it monotone} zero-finding problems. This means that each iteration can be solved efficiently by using a fixed point iteration (c.f. \cite{Ryu2020}, Chapter 2).

\subsection{Local Convergence of Mixed-Monotone Zero Finding Algorithm}
In Algorithm  \ref{algo_mixed_inclusion}, the mixed monotone update rule, which solves $0 \in A(x_{i+1}) -B(x_i)$ can be rewritten as 
\begin{equation}
	\label{update_mixed}
	x_{i+1}\in A^{-1} B(x_{i})
	\end{equation}
Suppose that $A$ is $\alpha$-coercive and $B$ is $\beta$-cocoercive. Then $A^{-1}B$ is  $\frac{\alpha}{\beta}$-Lipschitz. If  $\frac{\alpha}{\beta} < 1$, the update rule is a contraction mapping, and the iteration
converges to a unique fixed point. Moreover, this fixed point is a solution to the original
problem. 

We can generalize the convergnce result to the case where the properties hold only locally, allowing the result to be
applicable to the problems with multiple solutions.

To this end, we use the following version of Banach fixed-point theorem.
\begin{lem}\label{lem_bfpt}
	Suppose $\mathcal{D}$ is a $T$-invariant, closed subset of a Hilbert space where $T: \mathcal{D}\rightarrow \mathcal{D}$ is a contraction map. Then $T$ admits a unique fixed-point $x^{*} \in \mathcal{D}$ (i.e. $T(x^{*}) = x^{*}$). Furthermore, starting with an arbitrary element $x_0 \in \mathcal{D}$ there exists a sequence $\{x_i\}$ given by $x_{i+1} = T(x_i)$ for $i\geq 1$ such that $x_i \rightarrow x^{*}$ when $i\rightarrow \infty$.
	\end{lem}

\begin{pf}
The invariance of $\mathcal{D}$ under $T$ ensures that the sequence of iterates remains in $\mathcal{D}$. Moreover, the closedness of $\mathcal{D}$ and the completeness of a Hilbert space mean that $\mathcal{D}$ is complete, therefore contains the limit of the Cauchy sequence. Using these properties, the proof is identical to the standard Banach fixed-point theorem (e.g. see \cite{Agarwal2018}). 
	\end{pf}

\begin{thm}(Local Convergence of Algorithm \ref{algo_mixed_inclusion})\label{thm_converge}
	
	Given a Hilbert space $\mathcal{H}$ let $A: \mathcal{H}\rightarrow \mathcal{H}$, $B: \mathcal{H}\rightarrow \mathcal{H}$  and $\mathcal{D}\subseteq \mathcal{H}$ such that
	
	\begin{enumerate}
		\item $\mathcal{D}$ is closed,
        \item $A^{-1}B(\mathcal{D}) \subseteq \mathcal{D}$,
        \item $A$ is $\alpha$-coercive on $\mathcal{D}$,
        \item $B$ is $\beta$-cocoercive on $\mathcal{D}$,
        \item $\frac{\alpha}{\beta} < 1$.
		\end{enumerate}
 Then $A^{-1}B$ admits a unique fixed-point $x^{*} \in \mathcal{D}$ (i.e. $A^{-1}B(x^{*}) = x^{*}$). Furthermore, starting with an arbitrary element $x_0 \in \mathcal{D}$ there exists a sequence $\{x_i\}$ given by $x_{i+1} = A^{-1}B(x_i)$ for $i\geq 1$ such that $x_i \rightarrow x^{*}$ when $i\rightarrow \infty$.
\end{thm}
\begin{pf}
	The proof follows directly from Lemma \ref{lem_bfpt}, noting that the conditions of Theorem \ref{thm_converge} ensure that $A^{-1}B$ is $\frac{\alpha}{\beta}$-Lipschitz on $\mathcal{D}$. 
	\end{pf}
\begin{rem}
According to Theorem \ref{thm_converge}, existence of the closed subset $\mathcal{D}\subseteq \mcl H$ guarentess local existence and uniqueness of a fixed point in the iterative scheme $0\in A(x_{i+1})-B(x_i)$, hence the existence and uniqueness of a $x^{\star}$ such that $0\in A(x^{\star})- B(x^{\star})$. However, $A-B$ may have none, one or more than one zeros, making the selection of $\mathcal{D}$ problem-specific. 
	\end{rem}
\subsection{Illustration: Double-Well Potential Function}\label{ref_doublewell}
	As an example, consider the minimization problem of the double-well potential function (also considered in \cite{Lipp})
	\begin{align}
		\label{static_dc}
	\arg	\min\limits_{x\in \mathbb{R}}\   \frac{x^4}{12}- \frac{x^2}{2}.
	\end{align}
According to the first-order optimality condition, its critical point can be determined by  solving the following mixed monotone zero finding problem.
	\begin{align}
		\label{static_prob}
 0\in  \frac{x^3}{3}- x.
	\end{align}
Apart from the trivial solution $x=0$, there are two distinct zeros that are located on the left and right half plane equidistant from the origin respectively.

Using Algorithm \ref{algo_mixed_inclusion}, at every iteration $i>0$ and given $x_i$, one has to determine $x_{i+1}$ by solving $0=\frac{x_{i+1}^3}{3} - x_i$ that yields $x_{i+1} = \sqrt[3]{3x_i}$. 

To make the iterative scheme $x_{i+1} = \sqrt[3]{3x_i}$ locally convergent, one has to satisfy the conditions of Theorem \ref{thm_converge} by choosing a closed subset $\mcl D $ from which the initial guess $x_0$ is selcted. Note that, cubic function is $\alpha$-coercive on $\R \setminus \{ 0\}$ where the value of $\alpha$ is the slope of its tangent and it varies based on the domain. On the other hand, linear function is $\beta$-cocoercive on entire $\R$ where the value of $\beta$ is the slope. Therefore, we select two individual closed intervals $\mcl D^{+} \subseteq \R^{+}$ and $\mcl D^{-} \subseteq \R^{-}$. The condition $\frac{\alpha}{\beta}<1$ is achieved by keeping the lower interval of $\mcl D^{+}$ and $\mcl D^{-}$ small.  

\begin{figure}[h!]
		\centering
		\includegraphics[width=\linewidth]{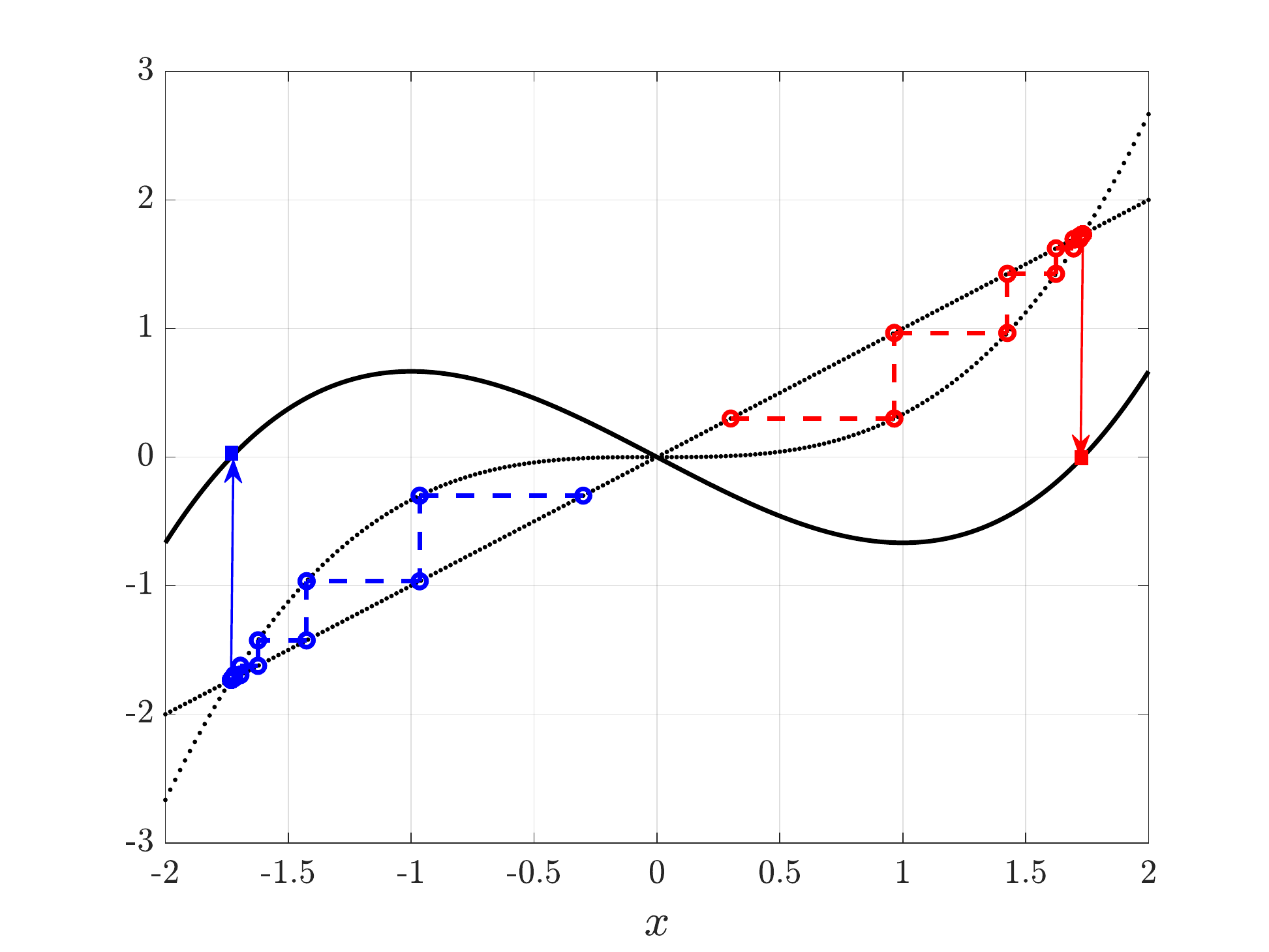}
		\caption{Geometrical enactment of Algorithm \ref{algo_mixed_inclusion} for finding zeros of $\frac{x^3}{3}- x$ (given in \crule{black}{1cm}{1pt}) that essentially computes the points of intersection between $y=x$ and $y=\frac{x^3}{3}$ (given in \protect\tikz[baseline]{\protect\draw[color=black,line width=0.75mm,densely dotted] (0,.8ex)--(1,.8ex)}). From two distinct intial guess $x_0 = 0.3, -0.3$, evolution of each iterate is depcited in \protect\tikz[baseline]{\protect\draw [color=red,  line width=0.55mm] (1.00, 0.00) circle (.05);\protect\draw [color=red,  line width=0.55mm] (2.00, 0.00) circle (.05);\protect\path [draw, dashed, color=red, line width=0.55mm] (1.05, 0.00) -- (1.95, 0.00)} and  \protect\tikz[baseline]{\protect\draw [color=blue,  line width=0.55mm] (1.00, 0.00) circle (.05);\protect\draw [color=blue,  line width=0.55mm] (2.00, 0.00) circle (.05);\protect\path [draw, dashed, color=blue, line width=0.55mm] (1.05, 0.00) -- (1.95, 0.00)} respectively.}		
		\label{fig:static_zero} 
	\end{figure}

Once intialized with $x_0 = 0$, the algorithm converges to the trivial solution $x = 0$. On the othr hand, based on whether the intial guess $x_0 $ is chosen to be $\mathcal{D}^{+}$ or $\mathcal{D}^{-}$, the algorithm always converges to either the positive or the negative fixed point respectively (in  this case, $x=\pm 1.7321$). Therefore, in case of mutiple solutions, Algorithm \ref{algo_mixed_inclusion} still converges, however, the convergence is local and depends on the intial guess. Grphical illustration of the algorithm is depicted in Figure \ref{fig:static_zero}. 

\tikzstyle{loose dashdotted}=  [dashed,mark=*,mark options={scale=2}]

\section{Finding a Periodic Solution to Mixed-Feedback System}
We aim to use the Algorithm \ref{algo_mixed_inclusion} to compute periodic solution of a mixed-feedback system as shown in figure \ref{fig:mixed_feed_sec2} casted as a mixed monotone zero finding problem. in other words, for a given input $u^{\star}\in l_{2, T}$,  we aim to compute the output $y^{\star}\in l_{2,T}$ such that 
\begin{align}
	\label{mm_1}
 &0\in  A(y^{\star})- B(y^{\star}),	\notag\\
\text{with,  } & A(y) := H^{-1} y + E_1(y)-u^{\star},\  B(y) :=  E_2(y).	
\end{align}
According to Alogorithm \ref{algo_mixed_inclusion}, finding $y^{\star}$ from \eqref{mm_1} requires to start from an intial guess $y_0$ that belongs to a preselected set $\mathcal{D} \subseteq l_{2,T}$ and iteratively compute the zero of the maximal monotone (locally) relation $A(y) -B(y_i)$ for every iteration $i>0$ untill the stopping criterion is satisfied.  

\subsection{Computing the zero of a locally monotone relation}
At every $i^{\text{th}}$ iteration, the maximal monotone relation $R(y;y_i):=A(y)-B(y_i)$ often admits a natural splitting such as $R= F_1+F_2$ where $F_1 $ and $F_2$ are maximal monotone operators. In such a case, the Douglas-Rachford splitting algorithm (c.f. \cite{Douglas1956}, \cite{Lions1979}) is applied, and the computation for the $F_1$ and
$F_2$ components are separated.  Given $y^{j}$, the Douglas-Rachford algorithm is given by the following iteration
\begin{align}
	\label{dr}
	w^{j + 1/2} =& \operatorname{res}_{F_1, \lambda}(y^j) \notag\\
	z^{j + 1/2} =& 2w^{j + 1/2} - y^j \notag\\
	w^{j+1} =& \operatorname{res}_{F_2, \lambda}(z^{j + 1/2}) \notag\\
	y^{j+1} =& y^j + w^{j+1} - w^{j + 1/2},
\end{align}
where $\operatorname{res}_{A, \lambda}$ denotes the resolvent of $A$, given by $(I + \lambda
A)^{-1}$, for parameter $\lambda > 0$.
The estimates $y$ and $w$ both converge to a fixed point of this iteration which is a
zero of $R=F_1+F_2$, as both $F_1$ and $F_2$ are maximal monotone.

\subsection{Complete Algorithm}
The complete algorithm employed to determine the solution to the mixed-monotone inclusion problem \eqref{mm_1} is as follows:

\begin{algorithm}[H]
	\caption{: Computing Solution to Mixed-Feedback Systems}
	\begin{algorithmic}[1]
		\State Given: $H,E_1, E_2, u^{\star}$ 
		\State Initialize: $i:=0$
		\State Given: An initial guess $y_0$, $\lambda$, $\epsilon_1$, $\epsilon_2$
		\For{$i=i+1$}
		\State Define $F_1 , F_2$ :  $$ \text{ such that } R= F_2+F_2 \text{ and } R(y, y_i):=A(y)-B(y_i)$$
		
		\State $\quad$ Initialize $j=0$, given an initial guess $y^0 = y_i$
		\State $\quad$ \textbf{for} $j=j+1$  \textbf{do}
		\begin{align*}
			w^{j + 1/2} =& \operatorname{res}_{F_1, \lambda}(y^j)\\
			z^{j + 1/2} =& 2w^{j + 1/2} - y^j\\
			w^{j+1} =& \operatorname{res}_{F_2,\lambda}(z^{j + 1/2})\\
			y^{j+1} =& y^j + w^{j+1} - w^{j + 1/2}
		\end{align*}
		\State $\quad$ \textbf{end for} when
		$\frac{\max \mid y^{j+1}-y^j \mid}{\max\mid y^j\mid}< \epsilon_2$ with $$y_{\star}=y^{j+1}$$ 
		\State $\quad$ \textbf{return} $y_{\star}$
		
		\State Update iteratively: $$y_{i+1} = y_{\star}$$
		\EndFor \hspace{1ex} when $\frac{\max \mid y_{i+1}-y_i \mid}{\max\mid y_i\mid}< \epsilon_1$ with $y^{\star}=y_{i+1}$ 
		\State \textbf{return} $y^{\star}$
	\end{algorithmic}
	\label{algo_total}
\end{algorithm}

\section{Illustration: Van Der Pol Oscillator}
The differential equation governing the Van der Pol Oscillator is 
\begin{equation}
	\label{vdr_pol}
	\ddot{x} + K (x^2-1)\dot{x} +x = 0,
\end{equation}
where, for time $t$, $x(t)\in \mathbb{R}$ is the state-variable and $K \geq 0$ is a constant parameter. The limit cycle solution to \eqref{vdr_pol}  and how it changes as a function of the parameter $K$ is well understood. In particular, when $K\rightarrow 0$, the limit cycle solution approaches a  pure sinusoid with frequency $1 \text{rad}/\text{sec}$. This limiting case is well captured by harmonic analysis. On the other hand, when $K \rightarrow \infty$, the limit cycle solution  approaches a  square-wave solution typical of {\it relaxation} oscillators \cite{vdr}. It is well known that the prediction of the describing function analysis is a periodic solution that has a natural frequency of $1  \text{rad}/\text{sec}$ and amplitude of $2$, regardless of the value of $K$ (see \cite{slotine}, Chapter 5). As a result, the describing function analysis fails to capture the change in the periodic solution for larger values of $K$. 

Here, the periodic solutions to the Van der Pol oscillator model is determined by the Algorithm \ref{algo_total} and involves the following considerations.

\subsubsection{Reformulation as a Mixed-Feedback System} 
\begin{figure}[h!]
	\centering
	\includegraphics[width=0.8\linewidth]{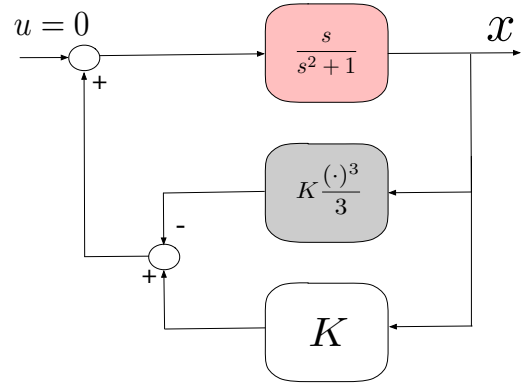}
	\caption{The mixed-feeback block diagram of the Van der Pol oscillator model. Here, $s(\cdot):= \frac{\text{d}}{\text{dt}}(\cdot)$}
	\label{fig:mixed_feed_vdr}
\end{figure}

The decomposition of Van der Pol model as a mixed feedback system is shown in   Figure \ref{fig:mixed_feed_vdr}. The positive feedback is naturally identified as the negative damping term $-K \dot x$ in (\ref{vdr_pol}).

As a result, computing a periodic solutio to the Van der Pol oscillator model amounts to computing the output $x^{\star}\in l_{2,T}$ such that 
\begin{align}
	\label{mm_vdr}
	&0\in  A(x^{\star})- B(x^{\star}),	\notag\\
	\text{with,  } & A(x) := \Bigg( \frac{s^2+1}{s}\Bigg) x + K \frac{x^3}{3},\  B(x) :=  K x.	
\end{align}
\subsubsection{Computation of the Algorithm}
According to Algorithm \ref{algo_total}, one must iteratively find a zero of $R(x, x_i):=A(x)-B(x_i)$ starting from an initial guess $x_0$ for each $i>0$ till the stopping criterion is met. Here, $R(x, x_i)$ admits a natural splitting such that $R=F_1+F_2$ where
\begin{align}
	F_1(x) :=& \Bigg( \frac{s^2+1}{s}\Bigg) x,\notag\\
	F_2(x,x_i):=& K \frac{x^3}{3}- K x_i.
\end{align}
Hence, to compute a zero of $A(x)-B(x_i)$ the Douglas-Rachford splitting algorithm is applied. Starting from an intial guess $x^{0} = x_i$, for each $j>0$ the resolvent of $F_1$ and $F_2$ are computed at the point $x^j$ and \eqref{dr} is computed iteratively. Since $F_1(x)$ is linear in $x$, for a given $x^j$, the computation of $w^{j+1/2}= \operatorname{res}_{F_1, \lambda}(x^j)$ amounts to solve $w^{j+1/2}$ such that
\begin{align}
	\label{resolvent_sys}
 &\Bigg(\lambda s^2  + s +\lambda  \Bigg) w^{j+1/2} =  s x^j.
\end{align}
To numerically evaluate the differential relations, the central difference scheme is employed, where for a function $f(t)$, $[s(f)](t):=\frac{\text{d}}{\text{d}t}f(t)\approx \frac{f(t+h)-f(t-h)}{2h}$ and $[s^2(f)](t):=\frac{\text{d}^2}{\text{d}t^2}f(t)\approx \frac{f(x+h)-2f(x)+f(x-h)}{h^2}$ for a given $h>0$. Moreover, to retain the periodicity, two conditions, $f(0-h) = f(T)$ and $f(T+h) = f(0)$, are enforced. As a result, determining $w^{j+1/2}$ in \eqref{resolvent_sys} amounts to solving a set of linear equations. 

On the other hand, finding the resolvent of $F_2(x):= K \frac{x^3}{3}- K x_i$ amounts to numerically finding the root of a cubic polynomial function with coefficients $(- K x_i, 0, 0, K)$. 
\subsubsection{Simulation Setting}
Algorithm \ref{algo_total} is implemented to compute the periodic solution of  \eqref{vdr_pol} for three different values of $K\in\{0.0002, 1.5, 10\} $, capturing three distinct regimes of oscillatory behaviour. It is expected that with increasing values of $K$ the periodic solution will change from pure sinusoids to square-waves, leading to the behavior of a relaxation oscillator. 

To obtain a periodic signal from the simulation, for each value of $K\in\{0.0002, 1.5, 10\} $, the time-interval is chosen to be $[0, nT_K]$, $n=1, 2, \cdots$, where $T_K$ is the fundamental period of solution for a given value of $K$.  Since the accurate value of the period is unknown for higher value of $K$, a rough approximation of the period is provided using M. Cartwright's semi-empirical formula (c.f. \cite{Letellier}). The time-interval is partitioned in $N$ equidistant points and the distance between two adjacent time points is denoted by $h$ where $h = \frac{nT_K}{N}$. For all simulations, we use $n=1$ and $N=5000$. The remaining user-defined paramaters are given in Table \ref{table:nonlin}. Moreover, the choice of initial guess is motivated by the illustration of double-well potential function (in subsection \ref{ref_doublewell}) since the addition of linear operator preserves the coercivity and cocoercivity. 
\begin{table}[ht]
	\caption{User-defined parameters for simulation}
	\centering
	\begin{tabular}{|c |c |c |c|}
		\hline
		Case & $\lambda$& $\epsilon_1= \epsilon_2$ & Initial Guess \\ [0.5ex] 
		\hline
		$K=0.0002$&$0.05$&$0.01$&Ramp with slope 1 \\
		$K=1.5$&$0.05$&$0.01$&Ramp with slope 1 \\
		$K=10$&$0.01$&$0.01$&Ramp with slope 1 \\ [1ex]
		\hline
	\end{tabular}
	\label{table:nonlin}
\end{table}
\subsubsection{Simulation Results}
\begin{figure}[h!]
	\centering
	\includegraphics[width=\linewidth]{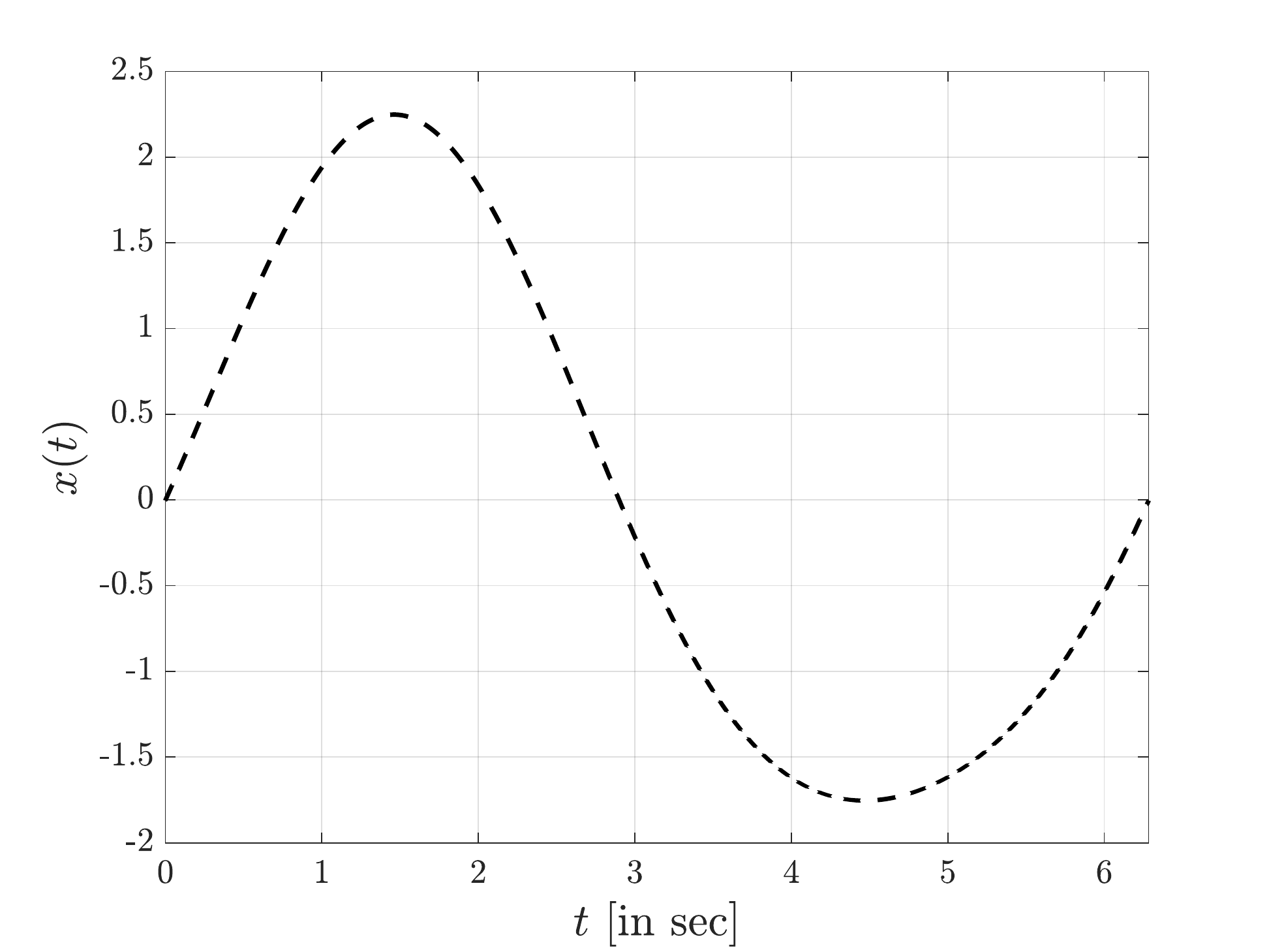}
	\caption{Periodic solution to the Van der Pol Oscillator for $K=0.0002$. The required number of iteration $i$ is $10$.}
	\label{fig:tau0_0002}
\end{figure}

\begin{figure}[h!]
	\centering
	\includegraphics[width=\linewidth]{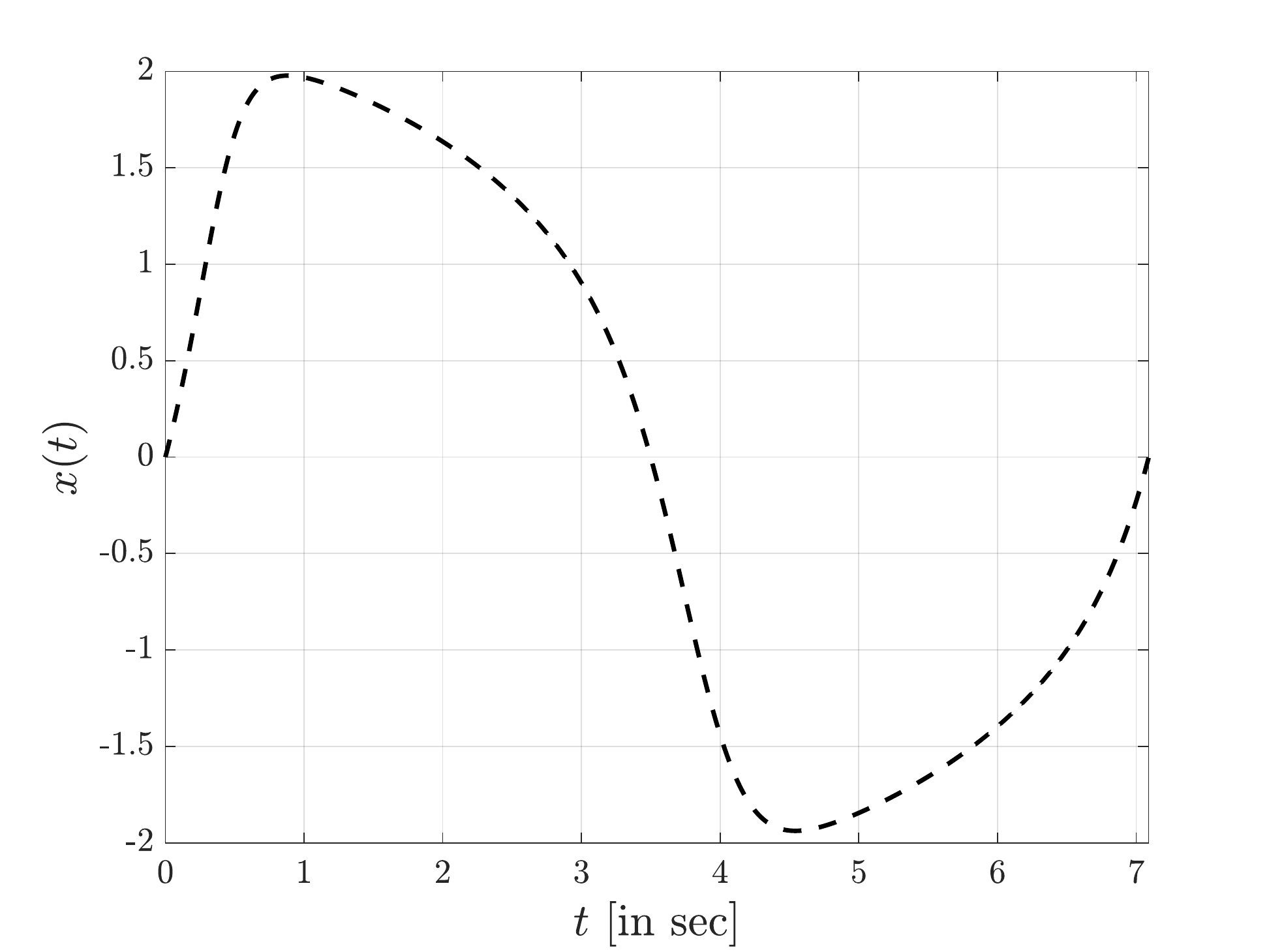}
	\caption{Periodic solution to the Van der Pol Oscillator for $K=1.5$. The required number of iteration $i$ is $8$.}
	\label{fig:tau1_5}
\end{figure}

\begin{figure}[h!]
	\centering
	\includegraphics[width=\linewidth]{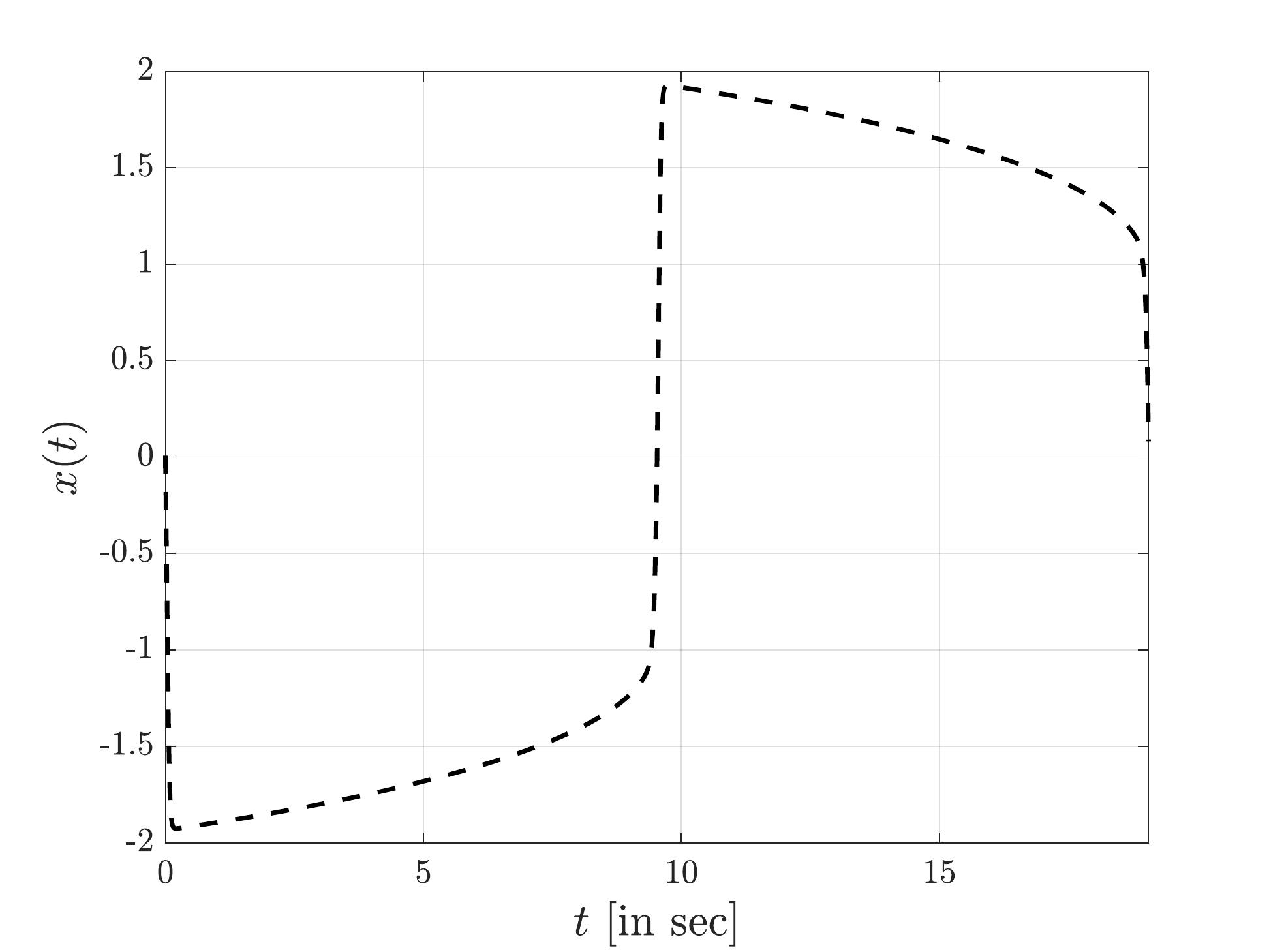}
	\caption{Periodic solution to the Van der Pol Oscillator for $K=10$. The required number of iteration $i$ is $8$.}
	\label{fig:tau10}
\end{figure}

The outputs of the algorithm provide the periodic solutions of \eqref{vdr_pol} and are depicted in figure \ref{fig:tau0_0002}-\ref{fig:tau10} along with the required number of iterations till the solution converges. As expected, the algorithm demonstrates the effective change in the oscillatory solution of the Van der Pol osciollator as the value of $K$ in \eqref{vdr_pol} increases from $0.0002$ till $10$. Evidently, the amplitude and period of the solution changes with different values of $K$, which is a significant distinction from the conclusion drawn from the describing function analysis for the Van der Pol oscillator. 

\section{Discussion and Future Work}

This paper borrows an effective algorithm of DC programing  to compute the limit cycle periodic solution of a general class of mixed-feedback systems.  Solutions of the feedback system are formulated as zeros of an relation. The mixed nature of the feedback system translates into the mixed monotonicity of the relation. For a given input, computing the output trajectory is formulated as finding zero of the difference between two maximal monotone relations and an iterative algorithm is presented that solves a zero finding problem for a sequence of {\it monotone} relations. The application of the algorithm to the classical Van der Pol oscillator for different parameter regimes suggests the efficiency and generality of the proposed approach. 

Due to the natural splitting between the feedforward passive operator and the feedback operators, the presented algorithm is scalable for the passive operator of arbitrarily large order. For instance, it is directly applicable to all mixed-feedback models considered in \cite{SEPULCHRE2005809}, \cite{stan} regardless of the dimension of the LTI passive operator.  

As the operators are evaluated on the space $l_{2,T}$ of $T$-periodic, square-summable signals, the algorithm requires some knowledge about the period. However, the knowledge of the period does not need to be accurate. The period is easily adapted over the course of the algorithm based on the zero crossings of the iterates.

We anticipate that the proposed method is general and offers an attractive complement to describing function analysis. We also envision natural extensions to the spatio-temporal mixed feeedback systems
such as the  PDE version of the Fitzhug-Nagumo model \cite{MIRANDAVILLATORO2021104858}, or standard models of Lateral Inhibition \cite{amari1977}. Ultimately, we hope to exploit the algorithmic framework  of  this paper  for the design and analysis of  multi-scale mixed-feedback systems such as those encountered in neuroscience and  neuromorphic engineering (see \cite{Sepulchre2018} for more details).

\begin{ack}                               
	The research leading to these results has received funding from the European Research Council under the Advanced ERC Grant Agreement Switchlet no. 670645.
\end{ack}

\bibliographystyle{unsrt}        
\bibliography{./IEEEabrv,./IEEEexample,./orchestron,./AR18}



\end{document}